\documentclass{aip-cp}

\usepackage[numbers]{natbib}
\usepackage{rotating}
\usepackage{graphicx}
\usepackage{bm}
\usepackage{amssymb}
\usepackage{color}

\begin{document}

\title{
Hyperon-Nucleon Interaction from Lattice QCD 
at $\bm{(m_\pi,m_K)\approx(146,525)}$~MeV
}

\author[aff1,aff2]{
Hidekatsu Nemura
\corref{cor1}
}
\affil{for HAL QCD Collaboration}

\affil[aff1]{
 Research Center for Nuclear Physics, Osaka University, Osaka, 567-0047, Japan
}
\affil[aff2]{
 Theoretical Research Division, Nishina Center, RIKEN, 
 Saitama, 351-0198, Japan
}
\corresp[cor1]{Corresponding author: 
hidekatsu.nemura@rcnp.osaka-u.ac.jp
}

\maketitle

\begin{abstract}
 Comprehensive study of generalized baryon-baryon ($BB$) interaction including 
 strangeness is one of the important subject of nuclear physics. 
 In order to obtain a complete set of isospin-base baryon interactions, 
 we perform a large scale lattice QCD calculation with almost physical 
 quark masses corresponding to 
 $(m_\pi,m_K)\approx(146,525)$ MeV and large volume 
 $(La)^4=(96a)^4\approx$ (8.1 fm)$^4$. 
 A large number of Nambu-Bethe-Salpeter (NBS) correlation functions 
 from nucleon-nucleon ($NN$) to $\Xi\Xi$ are 
 calculated simultaneously. %
 In this contribution, 
 we focus on the strangeness $S=-1$ channels of the hyperon
 interactions by means of HAL QCD method. 
%
%
%
%
%
Three potentials 
 ((i)~the $^1S_0$ central, (ii)~the $^3S_1-^3D_1$ central, and 
 (iii)~$^3S_1-^3D_1$ tensor potentials) are presented for 
 four isospin components; 
 (1)~the $\Sigma N - \Sigma N$ (the isospin $I=3/2$) diagonal, 
 (2)~the $\Lambda N - \Lambda N$ diagonal, 
 (3)~the $\Lambda N \rightarrow \Sigma N$ transition, and 
 (4)~the $\Sigma N - \Sigma N$ ($I=1/2$) diagonal. %
%
 Scattering phase shifts for $\Sigma N$ $(I=3/2)$ system are presented. 
%
\end{abstract}

%
\section{Introduction}

Elucidation of generalized nuclear forces including strangeness 
based on the fundamental perspectives (i.e, based on degrees of freedom 
in terms of quarks and gluons)
is one of the most important tasks of contemporary nuclear physics. 
For the normal nuclear interaction without strangeness, 
high precision 
experimental data are available; 
the interaction is described by a phenomenological approach 
that can reproduce %
the phase shifts and the deuteron 
properties with high accuracy. 
In addition, by combining a phenomenological three-nucleon force 
the energy levels of light nuclei can also be reproduced. 
On the other hand, for the hyperon-nucleon ($YN$) and hyperon-hyperon ($YY$) 
interactions, 
precise information such as phase shift analysis 
is limited because the scattering experiment of $YN$ or $YY$ system 
is difficult 
due to the hyperon short life-time. %
Thus far, 
based on the accumulation of accurate measurement of energy levels of 
various light hypernuclei~\cite{Hashimoto:2006aw} 
together with theoretical many-body studies~\cite{Yamamoto:2010zzn,Gal:2016boi} 
in addition to the (limited) scattering observables, 
a study toward comprehensive understanding the strange nuclear forces 
has been made. 
However, 
our knowledge of the $YN$ and $YY$ interactions is still away from 
the level of our knowledge of the $NN$ interaction. 
For the $\Sigma N$ interaction, 
only a four-body $\Sigma$-hypernucleus ($^4_{\Sigma}$He) has been observed and 
a repulsive $\Sigma$-nucleus interaction is inferred from the recent 
experimental study~\cite{Noumi:2001tx}. 
Such quantitative understanding is useful to study properties of 
high dense nuclear matters such as inside neutron stars, 
where recent observations of massive neutron stars heavier than 
$2M_{\odot}$ %
might raise a 
puzzle 
for the 
equation of state (EOS). %
Furthermore, 
keenly 
understanding of general nuclear forces %
would be 
important 
due to 
the recent 
observation of a 
binary neutron star merger~\cite{GBM:2017lvd,TheLIGOScientific:2017qsa}. %

During the last decade 
a new lattice QCD approach to study hadron-hadron interactions 
has been proposed~\cite{Ishii:2006ec,Aoki:2009ji} 
and 
developed to enhance the accuracy~\cite{HALQCD:2012aa}. 
In this approach, we first measure the Nambu-Bethe-Salpeter (NBS) 
wave function by means of the lattice QCD approach 
and then the interhadron potential is obtained. 
The scattering observables (e.g., phase shifts) and the binding energies are 
calculated by utilizing the potential. %
Thus far many studies have been performed by HAL QCD Collaboration
for the various baryonic 
interactions~\cite{Inoue:2011ai,Aoki:2011gt,Aoki:2012tk,Inoue:2013nfe,Murano:2013xxa,Inoue:2014ipa,Etminan:2014tya,Sasaki:2015ifa,Yamada:2015cra,Iritani:2016jie,Iritani:2017rlk,Gongyo:2017fjb}. 
This approach is now called HAL QCD method. 

In recent %
years, 2+1 flavor lattice QCD calculations have 
been 
extensively %
performed. 
Flavor symmetry breaking is a 
major concern 
in the study of 
$BB$ interactions.  
Therefore 
this is an opportune time to study 
the $BB$ potentials 
by using $2+1$ flavor lattice QCD. 
It is advantageous to calculate a 
large number of NBS wave functions of various $BB$ channels 
simultaneously in a single lattice QCD calculation. 
In these circumstances 
we 
consider 
the following $52$ 
four-point correlation functions 
in order to study the complete set of $BB$ interactions 
in the isospin symmetric limit~\cite{Nemura:2014eta,Nemura:2015yha}.
(For the moment, we assume that 
the electromagnetic interaction is not taken into account 
in the present lattice calculation.)
%
%
{%
\begin{eqnarray}
&&
\!\!\!\!
\!\!\!\!\!\!\!\!
\langle pn\overline{pn}\rangle,~ 
\label{GeneralBB_NN}
\\
&&
\!\!\!\!\!\!\!\!\!\!\!\!\!\!\!\!
\begin{array}{lll}
\langle p\Lambda\overline{p\Lambda}\rangle, &  \langle p\Lambda\overline{\Sigma^{+}n}\rangle, &  \langle p\Lambda\overline{\Sigma^{0}p}\rangle, 
\\
\langle \Sigma^{+}n\overline{p\Lambda}\rangle, &  \langle \Sigma^{+}n\overline{\Sigma^{+}n}\rangle, &  \langle \Sigma^{+}n\overline{\Sigma^{0}p}\rangle, 
\\
\langle \Sigma^{0}p\overline{p\Lambda}\rangle, &  \langle \Sigma^{0}p\overline{\Sigma^{+}n}\rangle, &  \langle \Sigma^{0}p\overline{\Sigma^{0}p}\rangle, 
\end{array}
\label{GeneralBB_NL}
\\
&&
\!\!\!\!\!\!\!\!\!\!\!\!\!\!\!\!
\begin{array}{llllll}
\langle \Lambda\Lambda\overline{\Lambda\Lambda}\rangle, &\!\!\!\! \langle \Lambda\Lambda\overline{p\Xi^{-}}\rangle, &\!\!\!\! \langle \Lambda\Lambda\overline{n\Xi^{0}}\rangle, &\!\!\!\! \langle \Lambda\Lambda\overline{\Sigma^{+}\Sigma^{-}}\rangle, &\!\!\!\! \langle \Lambda\Lambda\overline{\Sigma^{0}\Sigma^{0}}\rangle, 
\\
\langle p\Xi^{-}\overline{\Lambda\Lambda}\rangle, &\!\!\!\! \langle p\Xi^{-}\overline{p\Xi^{-}}\rangle, &\!\!\!\! \langle p\Xi^{-}\overline{n\Xi^{0}}\rangle, &\!\!\!\! \langle p\Xi^{-}\overline{\Sigma^{+}\Sigma^{-}}\rangle, &\!\!\!\! \langle p\Xi^{-}\overline{\Sigma^{0}\Sigma^{0}}\rangle, &\!\!\!\! \langle p\Xi^{-}\overline{\Sigma^{0}\Lambda}\rangle,~
\\
\langle n\Xi^{0}\overline{\Lambda\Lambda}\rangle, &\!\!\!\! \langle n\Xi^{0}\overline{p\Xi^{-}}\rangle, &\!\!\!\! \langle n\Xi^{0}\overline{n\Xi^{0}}\rangle, &\!\!\!\! \langle n\Xi^{0}\overline{\Sigma^{+}\Sigma^{-}}\rangle, &\!\!\!\! \langle n\Xi^{0}\overline{\Sigma^{0}\Sigma^{0}}\rangle, &\!\!\!\! \langle n\Xi^{0}\overline{\Sigma^{0}\Lambda}\rangle,~
\\
\langle \Sigma^{+}\Sigma^{-}\overline{\Lambda\Lambda}\rangle, &\!\!\!\! \langle \Sigma^{+}\Sigma^{-}\overline{p\Xi^{-}}\rangle, &\!\!\!\! \langle \Sigma^{+}\Sigma^{-}\overline{n\Xi^{0}}\rangle, &\!\!\!\! \langle \Sigma^{+}\Sigma^{-}\overline{\Sigma^{+}\Sigma^{-}}\rangle, &\!\!\!\! \langle \Sigma^{+}\Sigma^{-}\overline{\Sigma^{0}\Sigma^{0}}\rangle, &\!\!\!\! \langle \Sigma^{+}\Sigma^{-}\overline{\Sigma^{0}\Lambda}\rangle,~
\\
\langle \Sigma^{0}\Sigma^{0}\overline{\Lambda\Lambda}\rangle, &\!\!\!\! \langle \Sigma^{0}\Sigma^{0}\overline{p\Xi^{-}}\rangle, &\!\!\!\! \langle \Sigma^{0}\Sigma^{0}\overline{n\Xi^{0}}\rangle, &\!\!\!\! \langle \Sigma^{0}\Sigma^{0}\overline{\Sigma^{+}\Sigma^{-}}\rangle, &\!\!\!\! \langle \Sigma^{0}\Sigma^{0}\overline{\Sigma^{0}\Sigma^{0}}\rangle,~
\\
&\!\!\!\! \langle \Sigma^{0}\Lambda\overline{p\Xi^{-}}\rangle, &\!\!\!\! \langle \Sigma^{0}\Lambda\overline{n\Xi^{0}}\rangle, &\!\!\!\! \langle \Sigma^{0}\Lambda\overline{\Sigma^{+}\Sigma^{-}}\rangle, & &\!\!\!\! \langle \Sigma^{0}\Lambda\overline{\Sigma^{0}\Lambda}\rangle, 
\end{array}
\label{GeneralBB_LL}
\\
&&
\!\!\!\!\!\!\!\!\!\!\!\!\!\!\!\!
\begin{array}{lll}
\langle \Xi^{-}\Lambda\overline{\Xi^{-}\Lambda}\rangle, & \langle \Xi^{-}\Lambda\overline{\Sigma^{-}\Xi^{0}}\rangle, & \langle \Xi^{-}\Lambda\overline{\Sigma^{0}\Xi^{-}}\rangle,~
\\
\langle \Sigma^{-}\Xi^{0}\overline{\Xi^{-}\Lambda}\rangle, & \langle \Sigma^{-}\Xi^{0}\overline{\Sigma^{-}\Xi^{0}}\rangle, & \langle \Sigma^{-}\Xi^{0}\overline{\Sigma^{0}\Xi^{-}}\rangle,~
\\
\langle \Sigma^{0}\Xi^{-}\overline{\Xi^{-}\Lambda}\rangle, & \langle \Sigma^{0}\Xi^{-}\overline{\Sigma^{-}\Xi^{0}}\rangle, & \langle \Sigma^{0}\Xi^{-}\overline{\Sigma^{0}\Xi^{-}}\rangle,~
\end{array}
\label{GeneralBB_XL}
\\
&&
\!\!\!\!
\!\!\!\!\!\!\!\!
\langle \Xi^{-}\Xi^{0}\overline{\Xi^{-}\Xi^{0}}\rangle.~
\label{GeneralBB_XX}
\end{eqnarray}
}
%

A large scale lattice QCD calculation~\cite{Ishikawa:2015rho} is 
now in progress~\cite{Nemura:2017vjc,Ishii:2018ddn,Doi:2017zov,Sasaki:2018mzh} %
to study the baryon 
interactions from $NN$ to $\Xi\Xi$ 
by measuring 
a large number of 
NBS wave functions 
from $2+1$ flavor lattice QCD %
by employing the 
almost physical quark masses corresponding to 
 $(m_\pi,m_K)\approx(146,525)$ MeV. %
See also Ref.~\cite{Gongyo:2017fjb} for a study of 
the $\Omega\Omega$ interaction.

The purpose of this report is to present 
our recent results of the 
$\Lambda N-\Sigma N$ systems (both isospin values, $I=1/2$ and $3/2$) 
using full QCD gauge configurations. 
A very preliminary study had been reported at HYP2015 
with small statistics of $\Lambda N-\Lambda N$ single channel 
data~\cite{Nemura:2016sty}. 
This report shows the latest results of the study, 
based on recent works reported in 
Refs.~\cite{Nemura:2014eta,Nemura:2015yha};
the %
$YN$ interactions in the strangeness $S=-1$ sector 
(i.e, $\Lambda N-\Lambda N$, $\Lambda N-\Sigma N$, and 
$\Sigma N-\Sigma N$ (both $I=1/2$ and $3/2$)) are studied 
at almost physical quark masses 
corresponding to ($m_{\pi}$,$m_{K}$)$\approx$(146,525)~MeV and 
large volume $(La)^4=(96a)^4\approx$ (8.1 fm)$^4$ 
with the lattice spacing $a\approx 0.085$fm.

\section{Outline of the HAL QCD method}

In order to study the baryon-baryon interactions, 
we first define the equal time NBS wave function 
in particle channel $\lambda=\{B_{1},B_{2}\}$ 
with Euclidean time $t$~\cite{Ishii:2006ec,Aoki:2009ji} 
\begin{equation}
 \begin{array}{c}
 \phi_{\lambda E}(\vec{r}) {\rm e}^{-E t} = 
 \sum_{\vec{X}}
 \left\langle 0
  \left|
   B_{1,\alpha}(\vec{X}+\vec{r},t)
   B_{2,\beta}(\vec{X},t)
  \right| B=2, E, S, I 
 \right\rangle,
 \end{array}
  \label{DefineNBSWF}
\end{equation}
where 
$B_{1,\alpha}(x)$ ($B_{2,\beta}(x)$) denotes the local interpolating field of 
baryon $B_{1}$ ($B_{2}$) 
with mass $m_{B_{1}}$ ($m_{B_{2}}$), 
and 
$E=\sqrt{k_{\lambda}^2+m_{B_{1}}^2}+\sqrt{k_{\lambda}^2+m_{B_{2}}^2}$ 
is the total energy 
in the center of mass system of a baryon number $B=2$, strangeness $S$, 
and isospin $I$ state. 
For $B_{1,\alpha}(x)$ and $B_{2,\beta}(x)$, 
we employ the local interpolating field of octet baryons 
in terms of up ($u_{a\alpha}(x)$), down ($d_{a\alpha}(x)$), 
and strange ($s_{a\alpha}(x)$) quark field operators 
given by 
%
\begin{eqnarray}
  &&
  p_{\alpha} \! = \! \varepsilon_{abc} \left(
			 u_a C\gamma_5 d_b
			\right) u_{c\alpha},\!
\qquad
  n_{\alpha} \! = \! - \varepsilon_{abc} \left(
			   u_a C\gamma_5 d_b
			  \right) d_{c\alpha},\!
 \label{BaryonOperatorsOctet_N}
  \\
  &&
  \Sigma^{+}_{\alpha} \! = \! - \varepsilon_{abc} \left(
				    u_a C\gamma_5 s_b
				   \right) u_{c\alpha},\!
  \qquad
  \Sigma^{-}_{\alpha} \! = \! - \varepsilon_{abc} \left(
				    d_a C\gamma_5 s_b
				   \right) d_{c\alpha},\!
  \qquad
  \Sigma^{0}_{\alpha} \! = \! {1\over\sqrt{2}} \left( {X_u}_{\alpha} \! - \! {X_d}_{\alpha} \right),\!
 \label{BaryonOperatorsOctet_S}
  \\
  &&
  \Xi^{-}_{\alpha} \! = \! - \varepsilon_{abc} \left(
                                 d_a C\gamma_5 s_b
                                \right) s_{c\alpha},\!
  \qquad
  \Xi^{0}_{\alpha} \! = \! \varepsilon_{abc} \left(
                               u_a C\gamma_5 s_b
                              \right) s_{c\alpha},\!
 \label{BaryonOperatorsOctet_X}
  \\
  &&
  \Lambda_{\alpha} \! = \! {1\over \sqrt{6}} \left( {X_u}_{\alpha} \! + \! {X_d}_{\alpha} \! - \! 2 {X_s}_{\alpha} \right),\!
 \label{BaryonOperatorsOctet_L}
\end{eqnarray}
where
\begin{equation}
  {X_u}_{\alpha} = \varepsilon_{abc} \left( d_a C\gamma_5 s_b \right) u_{c\alpha}, 
  \qquad
  {X_d}_{\alpha} = \varepsilon_{abc} \left( s_a C\gamma_5 u_b \right) d_{c\alpha},
  \qquad
  {X_s}_{\alpha} = \varepsilon_{abc} \left( u_a C\gamma_5 d_b \right) s_{c\alpha}.
 \label{BaryonOperatorsOctet_XXX}
\end{equation}
%
%
The greek letters ($\alpha$) represent Dirac spinor and 
the roman letters ($a,b,c$) are 
the indices for the color. 
For simplicity, we have suppressed the explicit spinor indices in parenthesis 
and spatial coordinates in 
Equations~(\ref{BaryonOperatorsOctet_N})-(\ref{BaryonOperatorsOctet_XXX}) and 
the renormalization factors in Eq.~(\ref{DefineNBSWF}). 
Based on a set of the NBS wave functions, we define a non-local 
potential 
%
\begin{equation}
\begin{array}{c}
\left(
   \frac{\nabla^2}{2\mu_{\lambda}} + \frac{k_{\lambda}^{2}}{2\mu_{\lambda}}
  \right)
  \delta_{\lambda \lambda^{\prime}}
  \phi_{\lambda^{\prime} E}(\vec{r}) = 
  \int d^3r^\prime\, U_{\lambda\lambda^{\prime}}(\vec{r},\vec{r^{\prime}}) 
  \phi_{\lambda^{\prime} E}(\vec{r^{\prime}})
\end{array}
\end{equation}
with the reduced mass 
$\mu_{\lambda}=m_{B_{1}}m_{B_{2}}/(m_{B_{1}}+m_{B_{2}})$. 
%

In lattice QCD calculations, 
we compute the 
four-point correlation function defined by~\cite{HALQCD:2012aa} 
\begin{eqnarray}
 {F}_{\alpha\beta,JM}^{\langle B_1B_2\overline{B_3B_4}\rangle}(\vec{r},t-t_0) 
 && = 
 \sum_{\vec{X}}
 \left\langle  0 
  \left|
   B_{1,\alpha}(\vec{X}+\vec{r},t)
   B_{2,\beta}(\vec{X},t)
   \overline{{\cal J}_{B_{3} B_{4}}^{(J,M)}(t_0)}
  \right|  0 
 \right\rangle,
\end{eqnarray}
where 
$\overline{{\cal J}_{B_3B_4}^{(J,M)}(t_0)}=
  \sum_{\alpha^\prime\beta^\prime}
  P_{\alpha^\prime\beta^\prime}^{(J,M)}
  \overline{B_{3,\alpha^\prime}(t_0)}
  \overline{B_{4,\beta^\prime}(t_0)}$
is a source operator that creates $B_3B_4$ 
states with the
total angular momentum $J,M$. 
The normalized four-point function 
can be expressed as
\begin{eqnarray}
 &&
  {R}_{\alpha\beta,JM}^{\langle B_1B_2\overline{B_3B_4}\rangle}(\vec{r},t-t_0) 
  =
  {\rm e}^{(m_{B_1}+m_{B_2})(t-t_0)} 
  {F}_{\alpha\beta,JM}^{\langle B_1B_2\overline{B_3B_4}\rangle}(\vec{r},t-t_0) 
  \nonumber
  \\
  \!\!\!\!&=&\!\!\!\!
   \sum_{n} A_{n}
   \sum_{\vec{X}}
   \left\langle 0
    \left|
     B_{1,\alpha}(\vec{X}+\vec{r},0)
     B_{2,\beta}(\vec{X},0)
    \right| E_{n} 
   \right\rangle
   {\rm e}^{-(E_{n}-m_{B_1}-m_{B_2})(t-t_0)}
   \!+\! O({\rm e}^{-(E_{\rm th}-m_{B_{1}}-m_{B_{2}})(t-t_{0})}),
\label{NORMALIZED4PT}
\end{eqnarray}
where $E_n$ ($|E_n\rangle$) is the eigen-energy (eigen-state)
of the six-quark system 
and 
$A_n = \sum_{\alpha^\prime\beta^\prime} P_{\alpha^\prime\beta^\prime}^{(JM)}$
$\langle E_n | \overline{B}_{4,\beta^\prime}
\overline{B}_{3,\alpha^\prime} | 0 \rangle$. 
Hereafter, the spin and angular momentum subscripts are suppressed 
for $F$ and $R$ for simplicity. 
At moderately large $t-t_0$ 
where the 
inelastic contribution 
above the pion production 
$O({\rm e}^{-(E_{\rm th}-m_{B_{1}}-m_{B_{2}})(t-t_{0})})=
O({\rm e}^{-m_{\pi}(t-t_{0})})$ 
becomes 
negligible, 
we can construct the non-local potential $U$ through 
$\left(
   \frac{\nabla^2}{2\mu_{\lambda}} + \frac{k_{\lambda}^{2}}{2\mu_{\lambda}}
  \right)
  \delta_{\lambda \lambda^{\prime}}
  F_{\lambda^{\prime}}(\vec{r}) = 
  \int d^3r^\prime\, U_{\lambda \lambda^{\prime}}(\vec{r},\vec{r^{\prime}}) 
  F_{\lambda^{\prime}}(\vec{r^{\prime}}).$ 
In lattice QCD calculations in a finite box, it is practical to use 
the velocity (derivative) expansion, 
$U_{\lambda \lambda^{\prime}}(\vec{r},\vec{r^{\prime}}) =
 V_{\lambda \lambda^{\prime}}(\vec{r},\vec{\nabla}_{r})
\delta^{3}(\vec{r} - \vec{r^{\prime}}).$ 
In the lowest few orders we have 
\begin{equation}
V(\vec{r},\vec{\nabla}_{r}) = 
V^{(0)}(r) + V^{(\sigma)}(r)\vec{\sigma}_{1} \cdot \vec{\sigma}_{2} + 
V^{(T)}(r) S_{12}
 + 
V^{(^{\ LS}_{ALS})}(r) \vec{L}\cdot (\vec{\sigma}_{1}\pm\vec{\sigma}_{2})
 + 
O(\nabla^{2}),
\end{equation}
where $r=|\vec{r}|$, $\vec{\sigma}_{i}$ are the Pauli matrices acting 
on the spin space of the $i$-th baryon, 
$S_{12}=3
(\vec{r}\cdot\vec{\sigma}_{1})
(\vec{r}\cdot\vec{\sigma}_{2})/r^{2}-
\vec{\sigma}_{1}\cdot
\vec{\sigma}_{2}$ is the tensor operator, and 
$\vec{L}=\vec{r}\times (-i \vec{\nabla})$ is the angular momentum operator. 
The first three-terms constitute the leading order (LO) potential while 
the fourth term corresponds to the next-to-leading order (NLO) potential. 
By taking the non-relativistic approximation, 
$E_{n} - m_{B_{1}} - m_{B_{2}} \simeq 
 {k_{\lambda,n}^{2} \over {2\mu_{\lambda}}} +
 O(k_{\lambda,n}^{4})$, 
and neglecting the $V_{\rm NLO}$ and the higher order terms, 
we obtain 
\begin{equation}
\left(\frac{\nabla^2}{2\mu_{\lambda}} -\frac{\partial}{\partial t}\right)
{ R}_{\lambda\varepsilon}(\vec r,t)
\simeq 
V^{\rm (LO)}_{\lambda \lambda^{\prime}}(\vec{r}) 
\theta_{\lambda \lambda^{\prime}}
{ R}_{\lambda^{\prime}\varepsilon}(\vec r,t), \qquad \mbox{with} \qquad
\theta_{\lambda \lambda^{\prime}}=
{\rm e}^{( m_{B_{1}}+m_{B_{2}}-m_{B_{1}^{\prime}}-m_{B_{2}^{\prime}})(t-t_0)}.
\end{equation}
Note that we have introduced a matrix form 
${ R}_{\lambda^{\prime}\varepsilon} =
 \{R_{\lambda^{\prime}\varepsilon_{0}}, R_{\lambda^{\prime}\varepsilon_{1}}\}$
with linearly independent NBS wave functions 
$R_{\lambda^{\prime}\varepsilon_{0}}$ and 
$R_{\lambda^{\prime}\varepsilon_{1}}$.
%
For the spin
singlet
state, we extract the 
central potential as 
\begin{equation}
V_{\lambda \lambda^{\prime}}^{(Central)}(r;J=0)=
(\theta_{\lambda \lambda^{\prime}})^{-1}
({ R}^{-1})_{\varepsilon^{\prime}\lambda^{\prime}}
\left({\nabla^2\over 2\mu_{\lambda}}-{\partial\over \partial t}\right)
{ R}_{\lambda\varepsilon^{\prime}}. 
\end{equation}
For the spin triplet state, 
the wave function 
is decomposed into 
the $S$- 
and 
$D$-wave components as 
%
\begin{eqnarray}
  &&
  R%
  (\vec{r};\ ^3S_1)={\cal P}R%
  (\vec{r};J=1)
   \equiv {1\over 24} \sum_{{\cal R}\in{ O}} {\cal R}
   R%
   (\vec{r};J=1),
   \\
  &&
  R%
  (\vec{r};\ ^3D_1)={\cal Q}R%
  (\vec{r};J=1)
   \equiv (1-{\cal P})R%
   (\vec{r};J=1).
\end{eqnarray}
%
Therefore, 
the Schr\"{o}dinger equation with the LO 
potentials for the spin triplet state becomes
\begin{equation}
 \left\{
 \begin{array}{c}
  {\cal P} \\
  {\cal Q}
 \end{array}
 \right\}
 \times
 \left\{
  V^{(0)}_{\lambda \lambda^{\prime}}(r)
  +V^{(\sigma)}_{\lambda \lambda^{\prime}}(r)
  +V^{(T)}_{\lambda \lambda^{\prime}}(r)S_{12}
 \right\}
 \theta_{\lambda \lambda^{\prime}}
 { R}_{\lambda^{\prime}\varepsilon}(\vec{r},t-t_0)
 =
 \left\{
 \begin{array}{c}
  {\cal P} \\
  {\cal Q}
 \end{array}
 \right\}
 \times
 \left\{
     {\nabla^2\over 2\mu_{\lambda}} 
     -{\partial \over \partial t}
 \right\}
 { R}_{\lambda\varepsilon}(\vec{r},t-t_0),
\end{equation}
from which
the 
central and tensor potentials, 
$V_{\lambda\lambda^{\prime}}^{(Central)}(r;J=0)=
(V^{(0)}(r)-3V^{(\sigma)}(r))_{\lambda\lambda^{\prime}}$ for $J=0$, 
$V_{\lambda\lambda^{\prime}}^{(Central)}(r;J=1)=
(V^{(0)}(r) +V^{(\sigma)}(r))_{\lambda\lambda^{\prime}}$, 
and $V_{\lambda\lambda^{\prime}}^{(Tensor)}(r)$ for $J=1$, can be
determined\footnote{
The potential is obtained from the NBS 
wave function at moderately large imaginary time; it would be 
$t-t_{0} \gg 1/m_{\pi} \sim 1.4$~fm. 
In addition, 
no single state saturation between the ground state 
and the excited states with respect to the relative motion, 
e.g., 
$t-t_{0} \gg (\Delta E)^{-1} = 
\left( (2\pi)^2/(2\mu (La)^2) \right)^{-1} \simeq 8.0$~fm, 
is required for the HAL QCD method~\cite{HALQCD:2012aa}. 
}. 
%

\section{
Lattice setup 
}

$2+1$ flavor gauge configurations 
are generated on a $96^4$ lattice by employing the RG improved 
(Iwasaki) gauge action at $\beta=1.82$ with the nonperturbatively $O(a)$ 
improved Wilson quark (clover) action at 
$(\kappa_{ud},\kappa_{s})=(0.126117,0.124790)$ with $c_{sw}=1.11$ and 
the 6-APE stout smeared links with the smearing parameter $\rho=0.1$. 
The lattice QCD's measurement is performed at 
almost the physical quark masses; 
see Ref.~\cite{Ishikawa:2015rho} for details of the generation of the gauge configuration 
which show that 
light meson masses are 
$(m_{\pi},m_{K})\approx(146,525)$~MeV. 
The physical volume is 
$(aL)^4\approx$(8.1fm)$^4$ with the lattice 
spacing $a\approx 0.085$fm. 
Wall quark source is employed with Coulomb gauge fixing. 
For spacial direction the periodic boundary condition is used whereas 
for temporal direction the Dirichlet boundary condition (DBC) is used. 
The source and the DBC are separated by $|t_{DBC}-t_{0}|=48$. 
Each gauge configuration is used four times by using the hypercubic 
SO$(4,\mathbb{Z})$ symmetry of $96^4$ lattice. 
In order to further increase (double) the statistics 
forward and backward propagation 
in time are combined by using the charge conjugation and time reversal 
symmetries. 
A simultaneous calculation of 
a large number of baryon-baryon 
correlation functions including the 
channels from $NN$ to $\Xi\Xi$ 
is %
proposed %
and a %
C++ program is implemented~\cite{Nemura:2014eta}. %
The other program based on unified contraction algorithm 
(UCA)~\cite{Doi:2012xd} 
is implemented after the above work and 
the thoroughgoing consistency check in the numerical outputs 
is performed 
between the UCA and 
the present algorithm~\cite{Nemura:2015yha}. 
In this report, 
96 wall sources 
are used for the 
{414}
gauge configurations 
at every 
{5}
trajectories. 
The number of statistics has doubled from Ref.~\cite{Nemura:2017vjc}. 
Statistical data are averaged with the bin size 
{46}. 
Jackknife method is used to estimate the statistical errors.

\section{Results}

\subsection{Effective masses from single baryons' correlation function}
%
%
\begin{figure}[t]
  \centering \leavevmode
  \includegraphics[width=0.4752\textwidth]{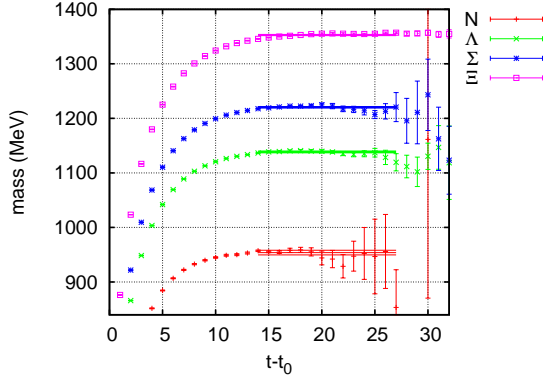}
  \caption{%
    The effective mass of single baryon's correlation functions 
    with utilizing wall sources. %
    \label{Fig_Effmass}}
\end{figure}
%
%
Figure~\ref{Fig_Effmass} shows the effective masses of 
the single baryon's correlation function. 
The plateaux start from time slices around $t-t_{0} \approx 14$
for the baryons $N,\Lambda$, and $\Sigma$. 
When calculating the normalized four-point correlation function in 
Eq.~(\ref{NORMALIZED4PT}). %
the %
exponential functional form ${\rm e}^{(m_{B_{1}}+m_{B_{2}})(t-t_{0})}$ 
is replaced by the single baryon's correlation functions, 
$(C_{B_{1}}(t-t_{0})C_{B_{2}}(t-t_{0}))^{-1}$. 
It would be beneficial to reduce the statistical noise 
because of 
the statistical correlation between the numerator and the denominator 
in the normalized four-point correlation function.
Therefore 
it is favorable that the potentials are obtained at 
the time slices $t-t_{0} \gtrsim 14$. 
In this report we present 
preliminary results of potentials 
at 
time slices ($t-t_0=5-14$) of our on-going work. 

\subsection{${\bm{ \Sigma N}}$ ($\bm{I=3/2}$) system}
\subsubsection{Potentials}
%
%
\begin{figure}[t]
 \begin{minipage}[t]{0.33\textwidth}
  \centering \leavevmode
%
  \includegraphics[width=0.99\textwidth]{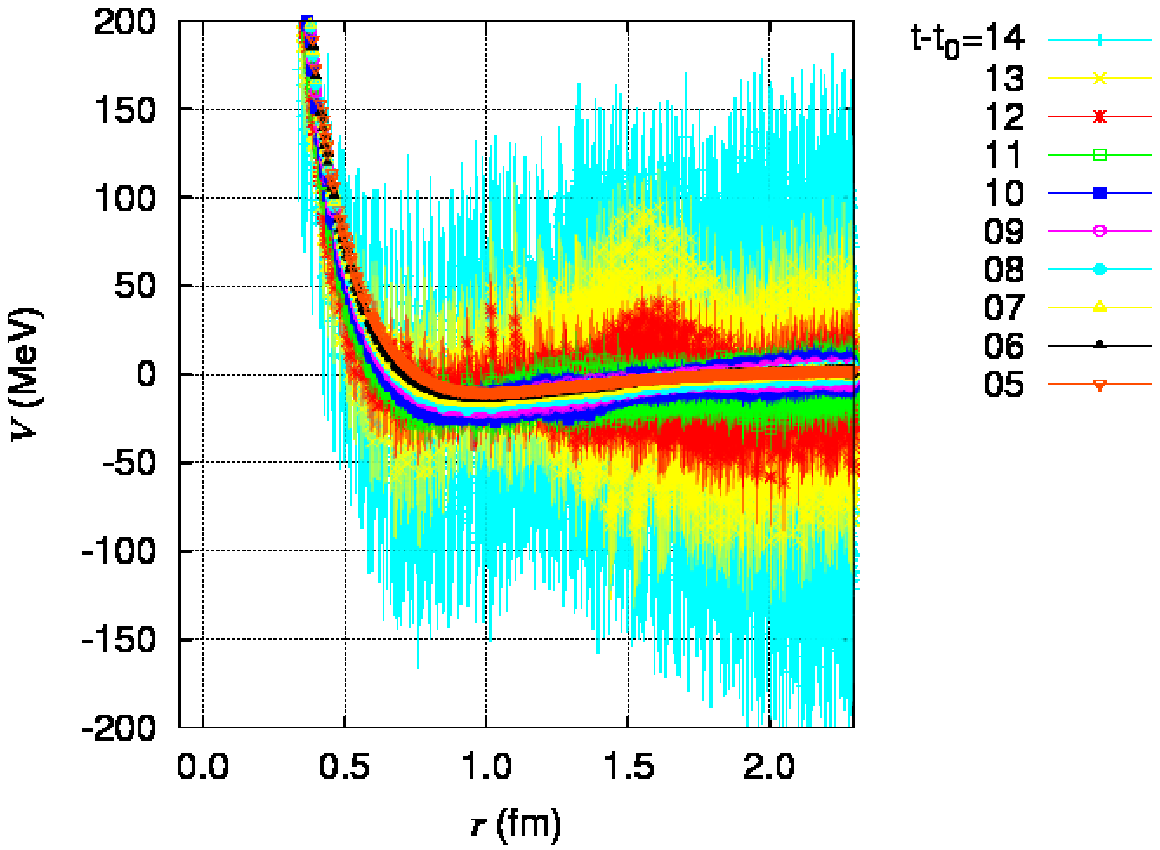}
%
 \end{minipage}~
 \hfill
 \begin{minipage}[t]{0.33\textwidth}
  \centering \leavevmode
%
  \includegraphics[width=0.99\textwidth]{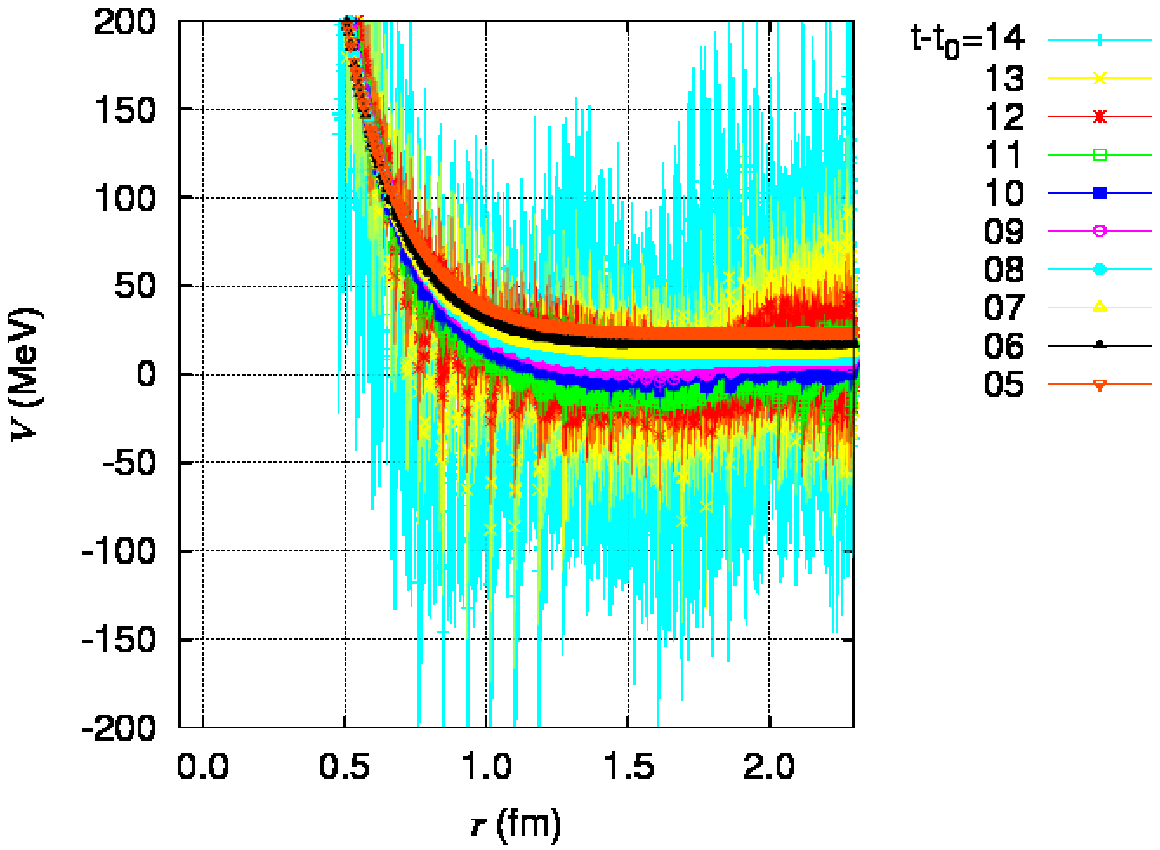}
%
 \end{minipage}~
 \hfill
 \begin{minipage}[t]{0.33\textwidth}
  \centering \leavevmode
%
  \includegraphics[width=0.99\textwidth]{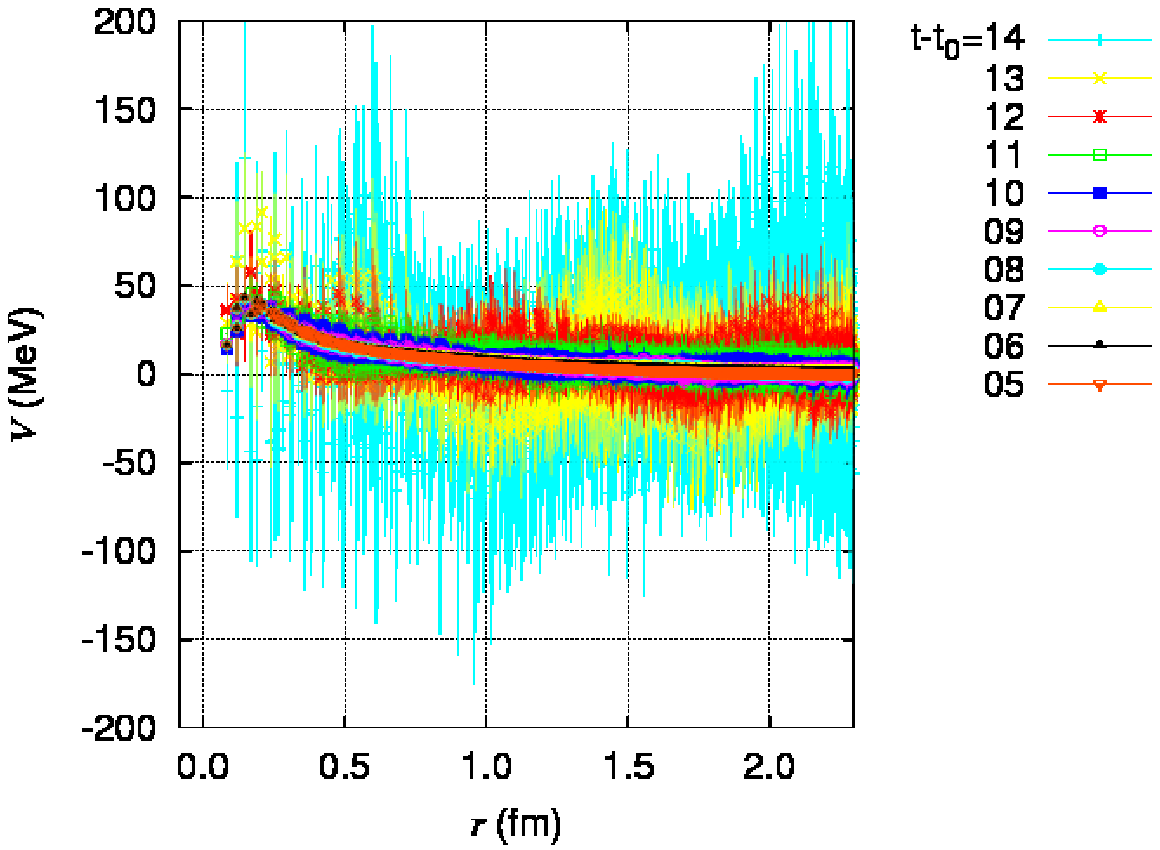}
%
 \end{minipage}
 \caption{Three $\Sigma N (I=3/2)$ potentials of 
   (i) $^1S_0$ central (left), 
   (ii) $^3S_1-^3D_1$ central (center), and 
   (iii) $^3S_1-^3D_1$ tensor (right). 
   \label{VC3E1_VT3E1_VC1S0_SN_2I3}}
\end{figure}
%
%
Fig.~\ref{VC3E1_VT3E1_VC1S0_SN_2I3} shows 
three potentials of $\Sigma N$ ($I=3/2$) system; 
(i)~the central potential in the $^1S_0$ (left), 
(ii)~the central potential in the $^3S_1-^3D_1$ (center), and 
(iii)~the tensor  potential in the $^3S_1-^3D_1$ (right). %
For the $^1S_0$ state 
there are both short ranged repulsive core and 
medium-to-long-distanced attractive well in the central potential. 
The potential is more or less similar to the $NN$ $^1S_0$ 
because this state belongs to flavor $\bm{27}$ 
irreducible representation ({\it irrep}). 
On the other hand, 
in the $^3S_1-^3D_1$ channel
a stronger repulsive core in the central potential is found. 
Especially the range of the repulsive core ($r \lesssim 1$~fm) 
is larger than the range of the repulsive 
core in the $^1S_0$ potential. 
This strong repulsive behavior is 
consistent with quark model's prediction 
that there is an almost Pauli forbidden state 
in the flavor $\bm{10}$ {\it irrep}. 
The tensor potential is not as strong as the $NN$ tensor potential. 
The statistical fluctuation of the tensor potential becomes large at the time 
slices $t-t_0 \ge 11$ whereas 
that of the tensor potential 
at $t-t_0 \le 10$ 
does not. 
These observations are consistent with the scattering phase shift 
calculated below. 

\subsubsection{Scattering phase shifts}

%
\begin{figure}[b]
  \centering \leavevmode
  \includegraphics[width=0.3267\textwidth]{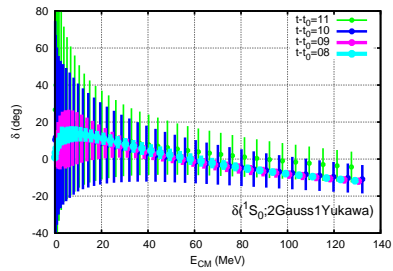}
 \footnotesize
 \caption{Scattering phase shift in the $^1S_0$ state 
   of $\Sigma N (I=3/2)$ system, 
   obtained by 
   solving the Schr\"{o}dinger equation with %
   parametrized functional form Eq.~(\ref{VCndVT}). 
   \label{Fig_Phsft1S0SN2I3}}
\end{figure}
\begin{figure}[t]
 \begin{minipage}[t]{0.33\textwidth}
  \centering \leavevmode
  \includegraphics[width=0.99\textwidth]{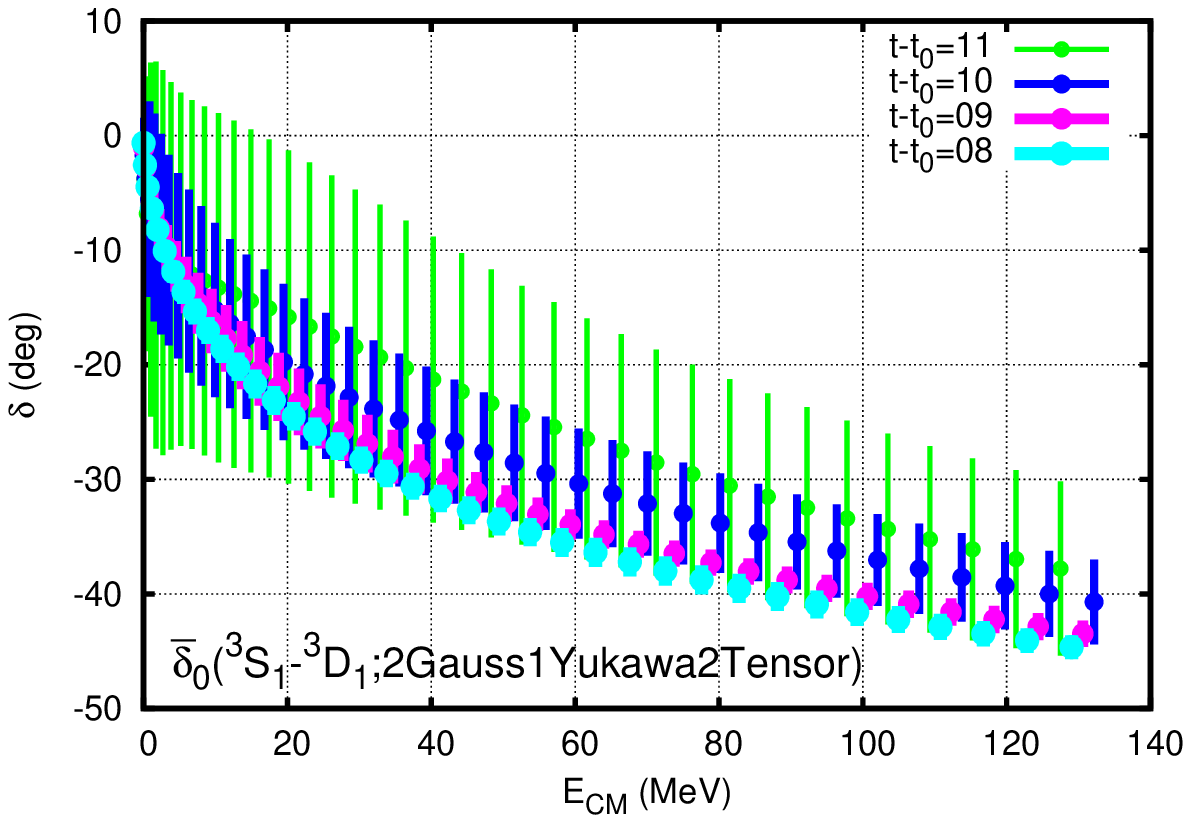}
 \end{minipage}~
 \hfill
 \begin{minipage}[t]{0.33\textwidth}
  \centering \leavevmode
  \includegraphics[width=0.99\textwidth]{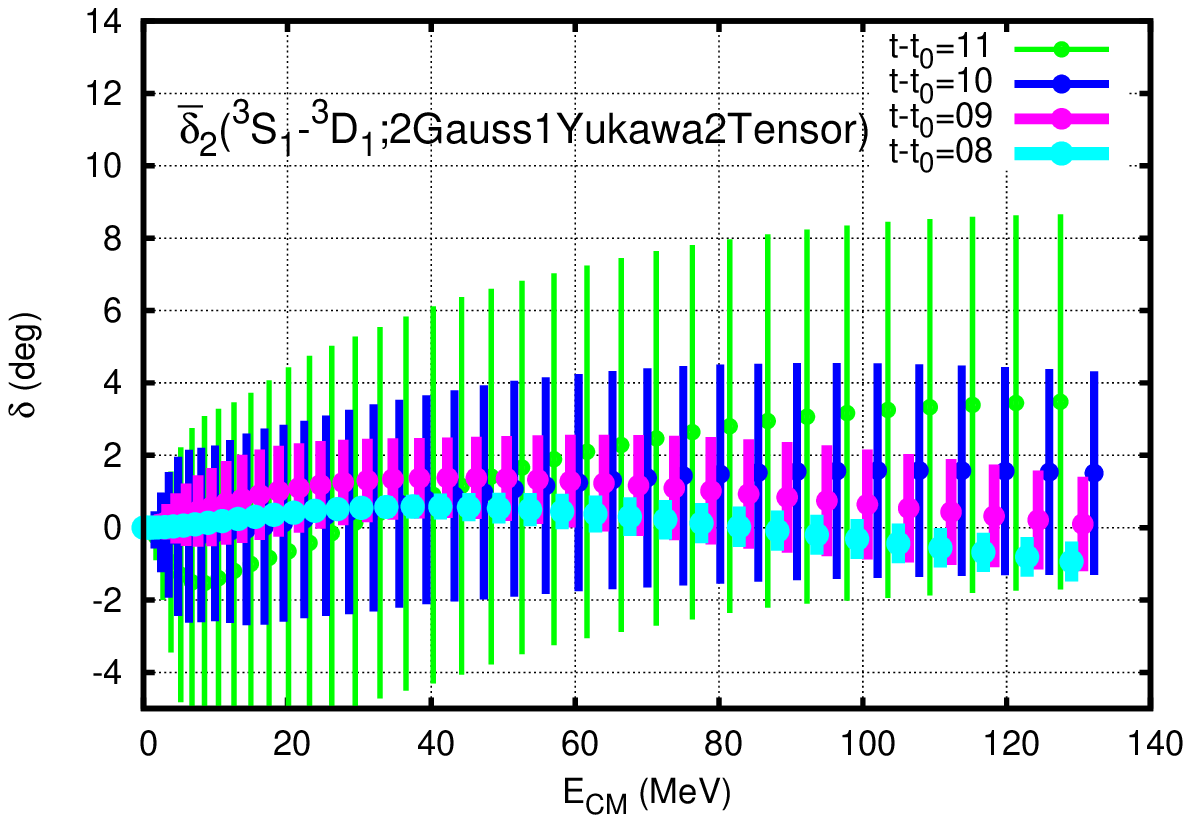}
 \end{minipage}~
 \hfill
 \begin{minipage}[t]{0.33\textwidth}
  \centering \leavevmode
  \includegraphics[width=0.99\textwidth]{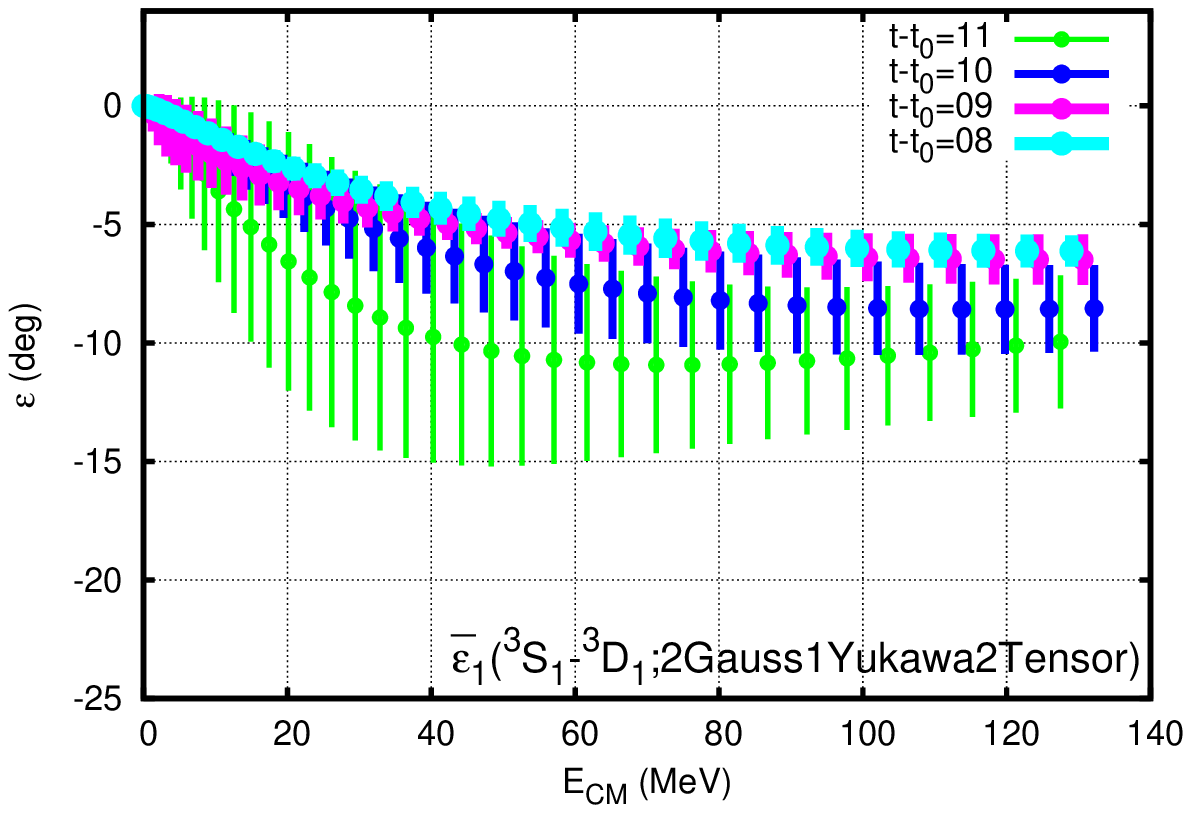}
 \end{minipage}
  \caption{Scattering bar-phase shifts and mixing angle in the 
    $^3S_1-^3D_1$ states of $\Sigma N (I=3/2)$ system, 
    $\bar{\delta}_0$ (left), 
    $\bar{\delta}_2$ (center), and 
    $\bar{\varepsilon}_1$ (right), 
    obtained by solving the Schr\"{o}dinger equation 
    with parametrized functional form Eq.~(\ref{VCndVT}).  
    \label{Fig_Phsft3E1SN2I3}}
\end{figure}

%
%
In order to obtain the scattering phase shift from 
the lattice QCD potential obtained above 
we first parametrize the potential %
with an analytic functional form. 
In this report, 
we use following functional forms for the central and 
tensor potentials, %
respectively. 
\begin{equation}
  \begin{array}{l}
    V_{C}(r)
    = v_{C1} {\rm e}^{-\kappa_{C1} r^2}
    + v_{C2} {\rm e}^{-\kappa_{C2} r^2}
    + v_{C3}\left( 1-{\rm e}^{-\alpha_{C} r^2}\right)^2 
    \left( {{\rm e}^{-\beta_{C} r}\over r} \right)^2, 
    \\
    V_{T}(r)
    = v_{T1} \left( 1-{\rm e}^{-\alpha_{T1} r^2} \right)^{2} 
    \left( 1+{3\over \beta_{T1} r}+{3\over (\beta_{T1} r)^2} \right) 
         {{\rm e}^{-\beta_{T1} r}\over r}
    + v_{T2} \left( 1-{\rm e}^{-\alpha_{T2} r^2} \right)^{2} 
    \left( 1+{3\over \beta_{T2} r}+{3\over (\beta_{T2} r)^2} \right) 
              {{\rm e}^{-\beta_{T2} r}\over r}.
  \end{array}
  \label{VCndVT}
\end{equation}
Figure~\ref{Fig_Phsft1S0SN2I3} shows the scattering phase shift in $^1S_0$ 
channel of $\Sigma N (I=3/2)$ system obtained through 
the above parametrized potentials. 
The present result shows that the interaction in the $^1S_0$ channel 
is attractive on average 
though the fluctuation is 
large especially for the time slices $t-t_0=10,11$. 
Figure~\ref{Fig_Phsft3E1SN2I3} shows the scattering phase shifts in 
$^3S_1-^3D_1$ channels. 
For the $^3S_1-^3D_1$ channels, 
the scattering matrix is parametrized with three real parameters 
bar-phase shifts and mixing angle: %
\begin{equation}
S = 
  \left( \begin{array}{cc}
    {\rm e}^{i \bar{\delta}_{J-1}} & 0 \\ %
    0 & {\rm e}^{i \bar{\delta}_{J+1}}    %
  \end{array} \right)
  \left( \begin{array}{cc}
      \cos 2 \bar{\varepsilon}_{J} & i \sin 2 \bar{\varepsilon}_{J} \\
    i \sin 2 \bar{\varepsilon}_{J} &   \cos 2 \bar{\varepsilon}_{J} 
  \end{array} \right)
  \left( \begin{array}{cc}
    {\rm e}^{i \bar{\delta}_{J-1}} & 0 \\ %
    0 & {\rm e}^{i \bar{\delta}_{J+1}}    %
  \end{array} \right).
\end{equation}
The phase shift $\bar{\delta}_0$ at the time slices $t-t_0=9-11$ shows 
the interaction is repulsive 
while the phase shift $\bar{\delta}_2$ behaves around almost zero degree. 
%
%
%
%

%
\subsection{$\bm{\Lambda N-\Sigma N}$ ($\bm{I=1/2}$) coupled-channel systems}
%
%
\begin{figure}[b]
 \begin{minipage}[t]{0.33\textwidth}
  \centering \leavevmode
%
  \includegraphics[width=0.99\textwidth]{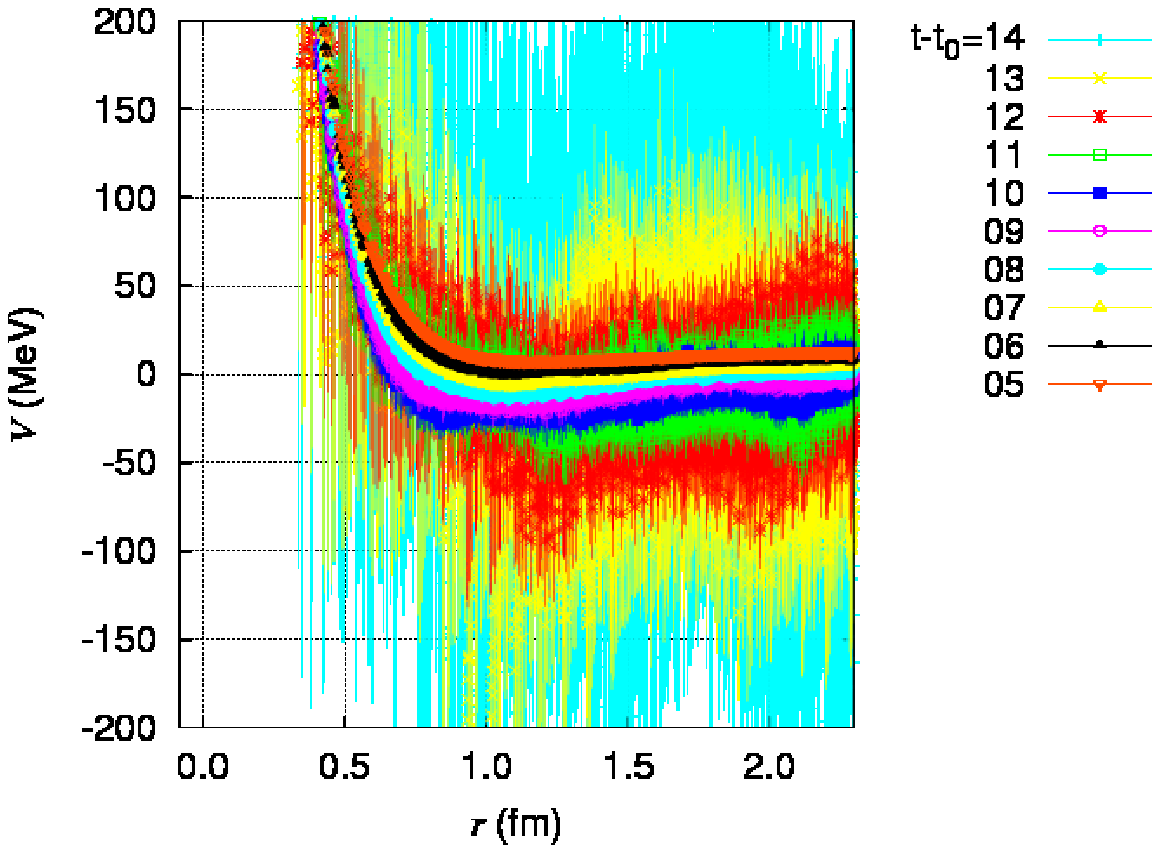}
%
 \end{minipage}~
 \hfill
 \begin{minipage}[t]{0.33\textwidth}
  \centering \leavevmode
%
  \includegraphics[width=0.99\textwidth]{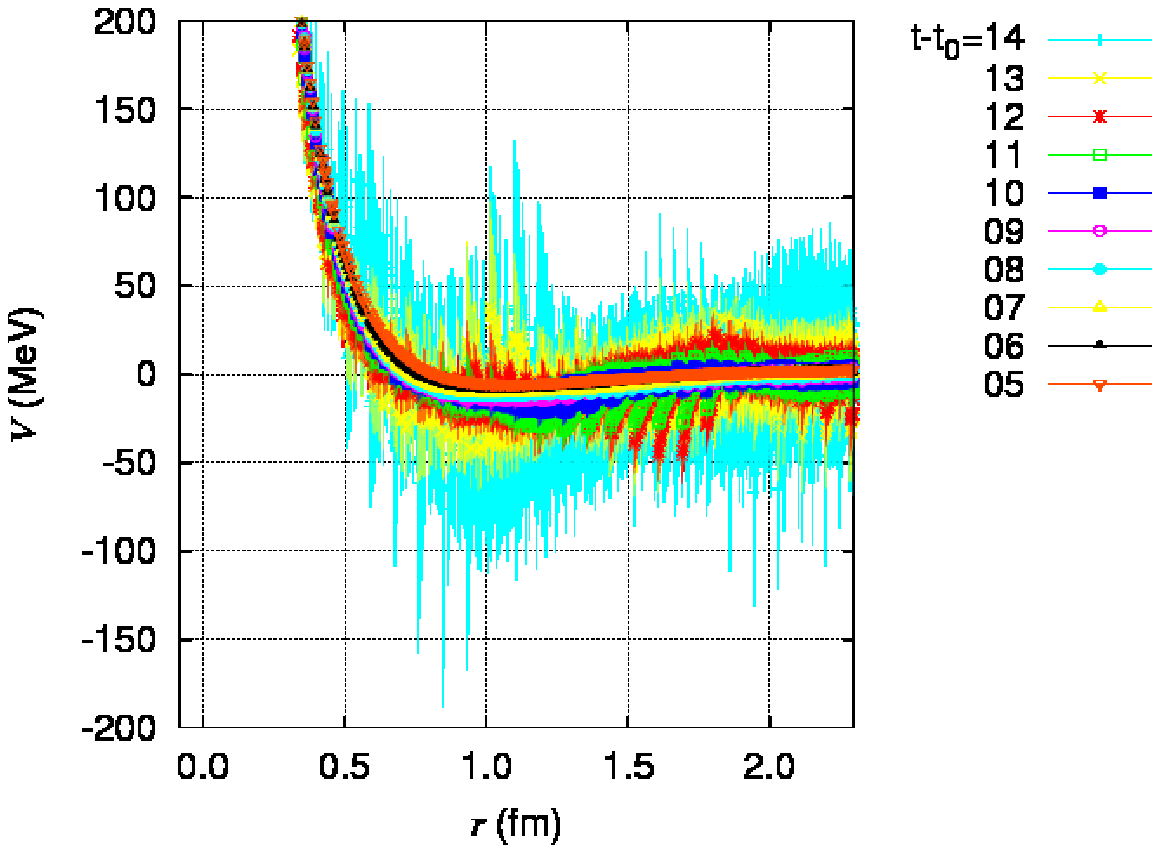}
%
 \end{minipage}~
 \hfill
 \begin{minipage}[t]{0.33\textwidth}
  \centering \leavevmode
%
  \includegraphics[width=0.99\textwidth]{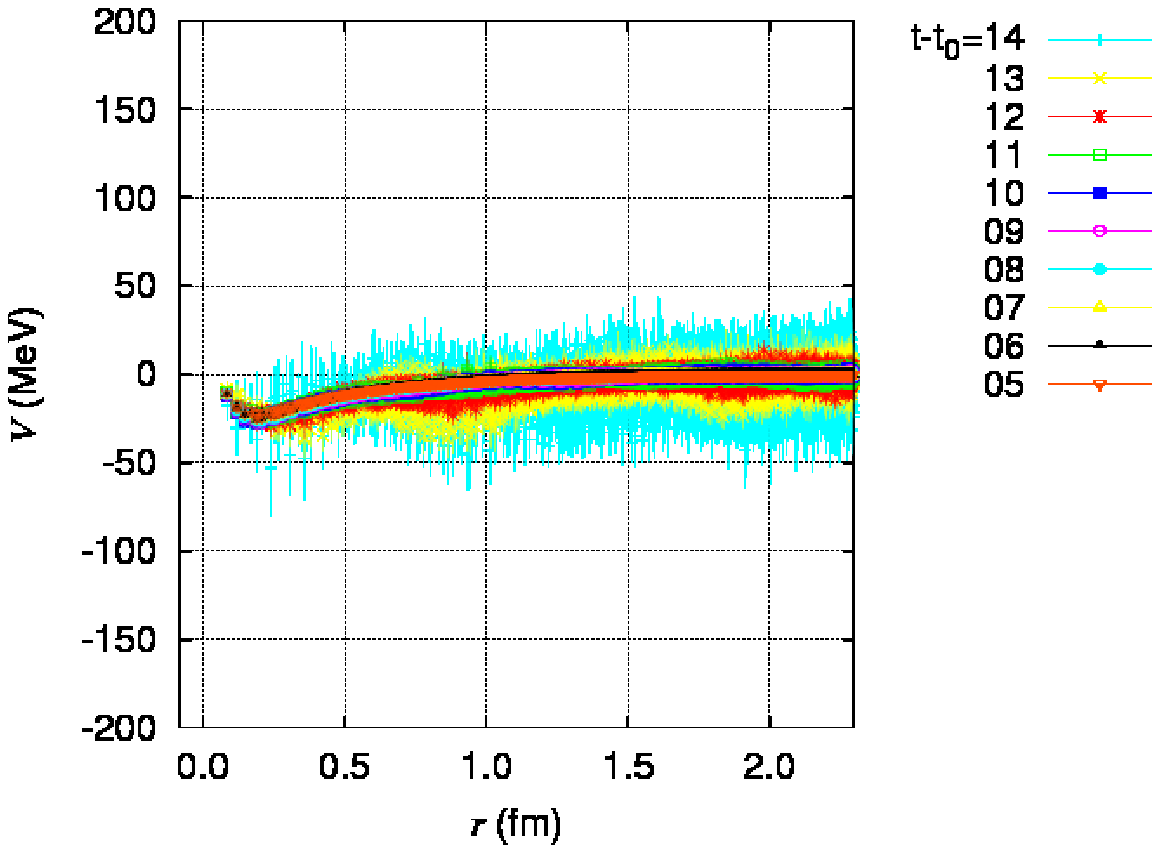}
%
 \end{minipage}
 \caption{Three $\Lambda N-\Lambda N$ potentials for 
   (i) $^1S_0$ central (left), 
   (ii) $^3S_1-^3D_1$ central (center), and 
   (iii) $^3S_1-^3D_1$ tensor (right). 
   \label{VC3E1_VT3E1_VC1S0_LN}}
\end{figure}
%
%
Fig.~\ref{VC3E1_VT3E1_VC1S0_LN} shows 
three $\Lambda N-\Lambda N$ diagonal potentials; 
(i) the central potential in the $^1S_0$ (left),
(ii) the central potential in the $^3S_1-^3D_1$ (center), and 
(iii) the tensor potential in the $^3S_1-^3D_1$ (right). %
There are repulsive cores in the short distance region and medium to long range 
attractive well for both central potentials. 
In the $\Lambda N-\Lambda N$ diagonal part, 
the tensor potential is relatively weak. 
%
%
\begin{figure}[t]
 \begin{minipage}[t]{0.33\textwidth}
  \centering \leavevmode
%
  \includegraphics[width=0.99\textwidth]{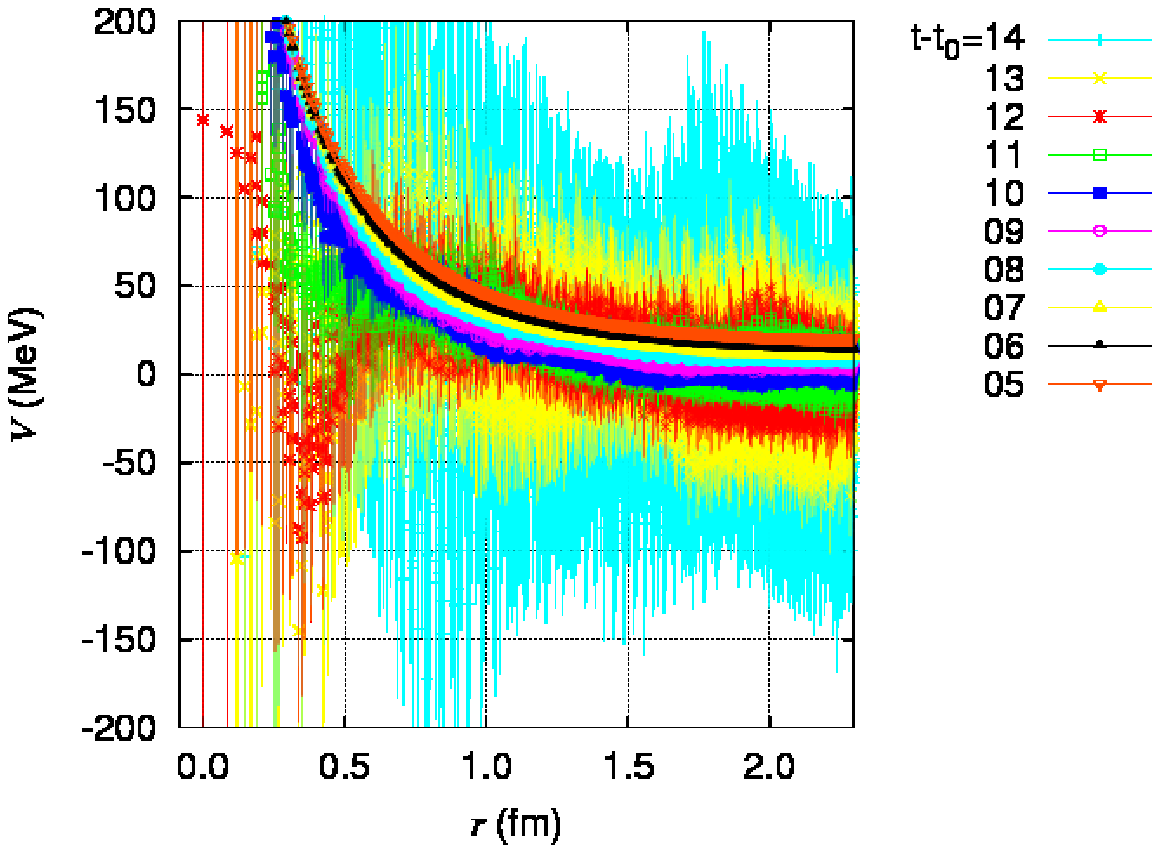}
%
 \end{minipage}~
 \hfill
 \begin{minipage}[t]{0.33\textwidth}
  \centering \leavevmode
%
  \includegraphics[width=0.99\textwidth]{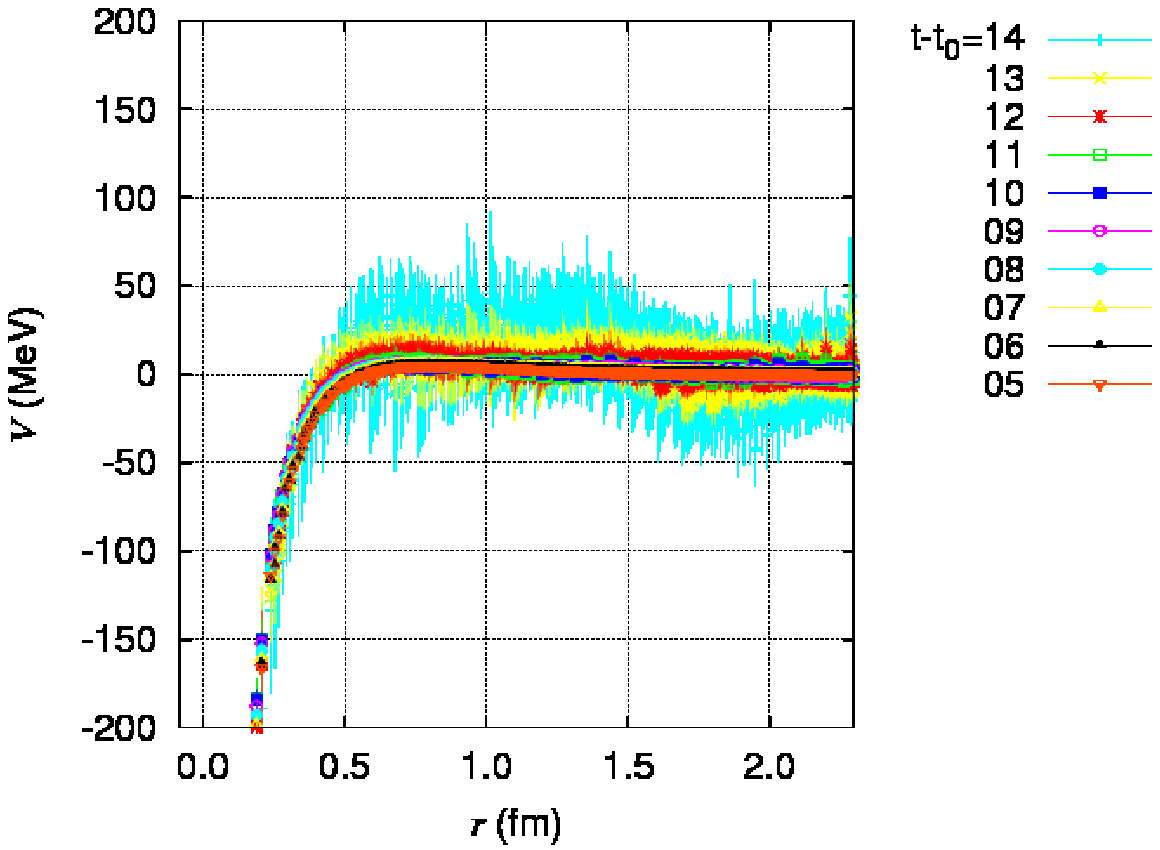}
%
 \end{minipage}~
 \hfill
 \begin{minipage}[t]{0.33\textwidth}
  \centering \leavevmode
%
  \includegraphics[width=0.99\textwidth]{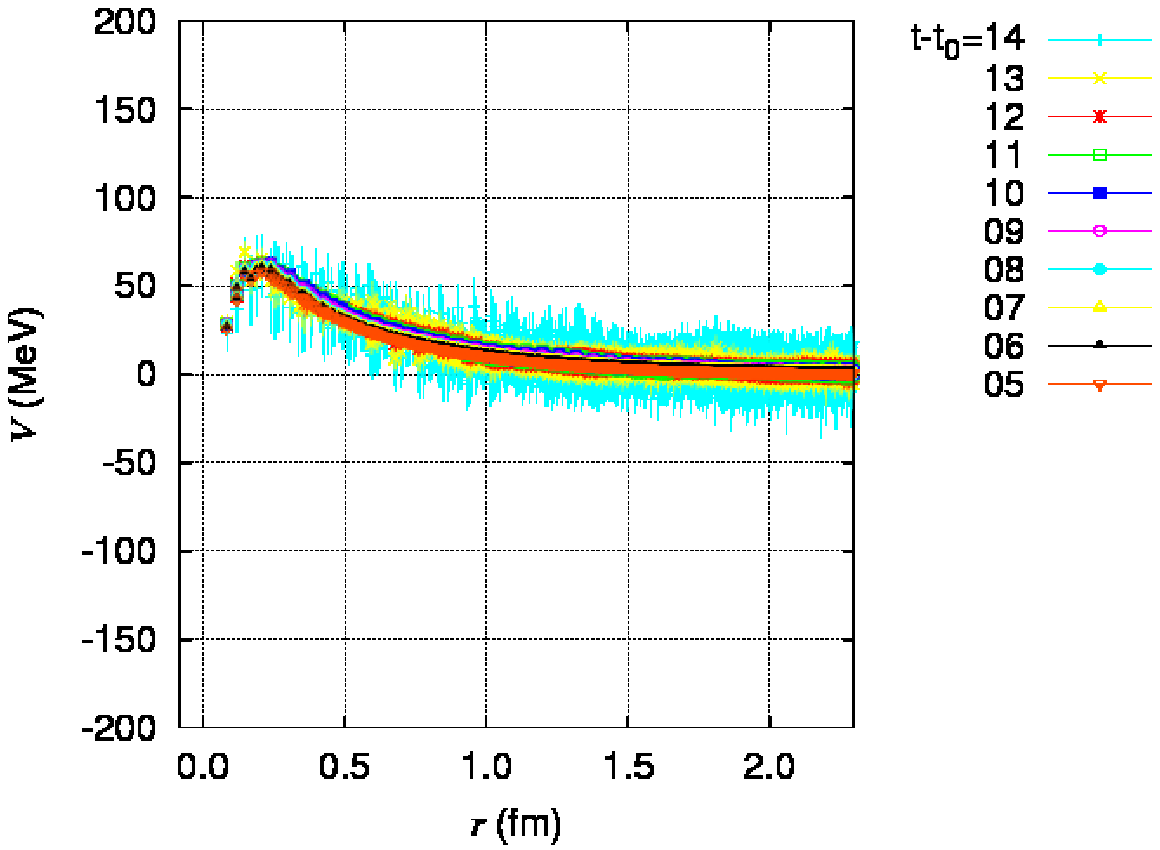}
%
 \end{minipage}
 \caption{Three $\Lambda N\rightarrow\Sigma N$ potentials for 
   (i) $^1S_0$ central (left), 
   (ii) $^3S_1-^3D_1$ central (center), and 
   (iii) $^3S_1-^3D_1$ tensor (right). 
   \label{VC3E1_VT3E1_VC1S0_LNtoSN}}
\end{figure}
%
%
Fig.~\ref{VC3E1_VT3E1_VC1S0_LNtoSN} shows 
three potentials of the 
$\Lambda N\rightarrow\Sigma N$ transition part; 
(i) the central potential in the $^1S_0$ (left),
(ii) the central potential in the $^3S_1-^3D_1$ (center), and 
(iii) the tensor  potential in the $^3S_1-^3D_1$ (right). %
The statistical fluctuation in the $^1S_0$ central potential is still large. 
The $^3S_1-^3D_1$ central potential is 
short ranged. 
In the $\Lambda N\rightarrow\Sigma N$ off-diagonal part, 
the tensor potential shows a sizable strength although 
it is not as strong as the $NN$ tensor potential. 
%
%
\begin{figure}[b]
 \begin{minipage}[t]{0.33\textwidth}
  \centering \leavevmode
%
  \includegraphics[width=0.99\textwidth]{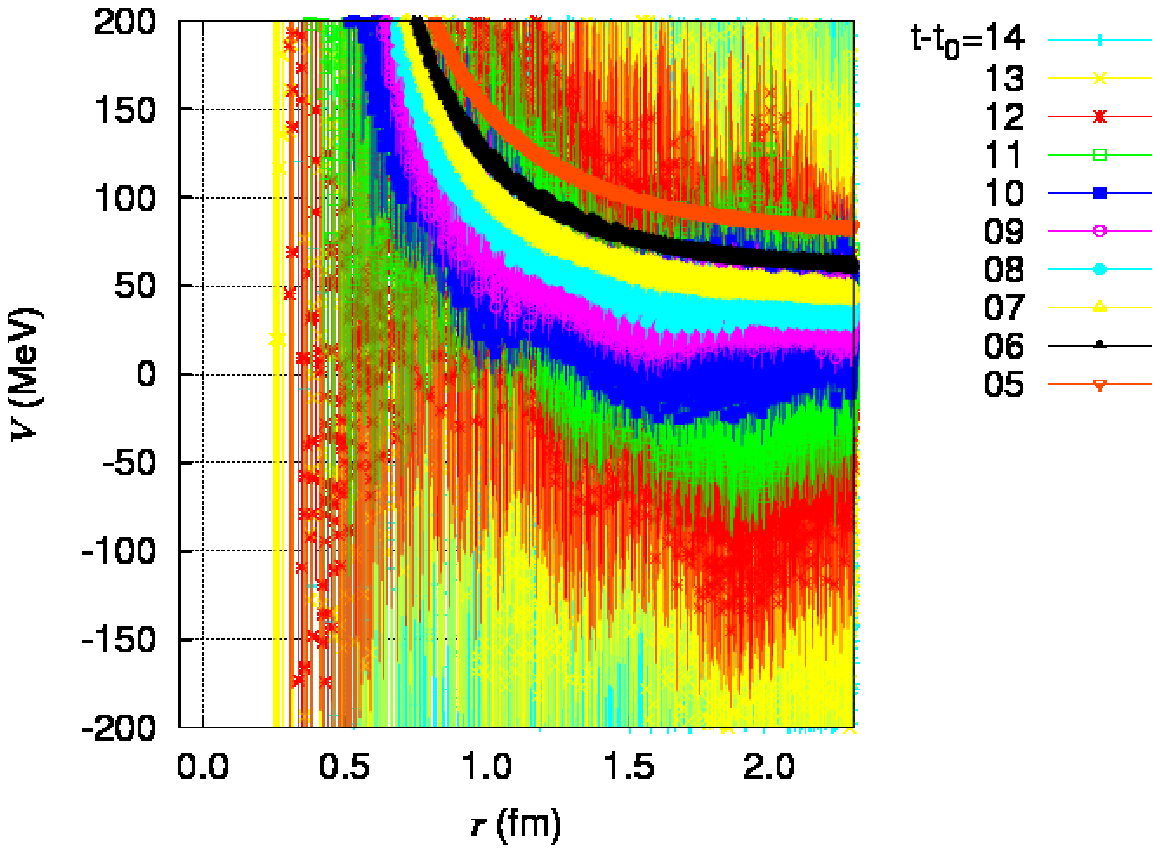}
%
 \end{minipage}~
 \hfill
 \begin{minipage}[t]{0.33\textwidth}
  \centering \leavevmode
%
  \includegraphics[width=0.99\textwidth]{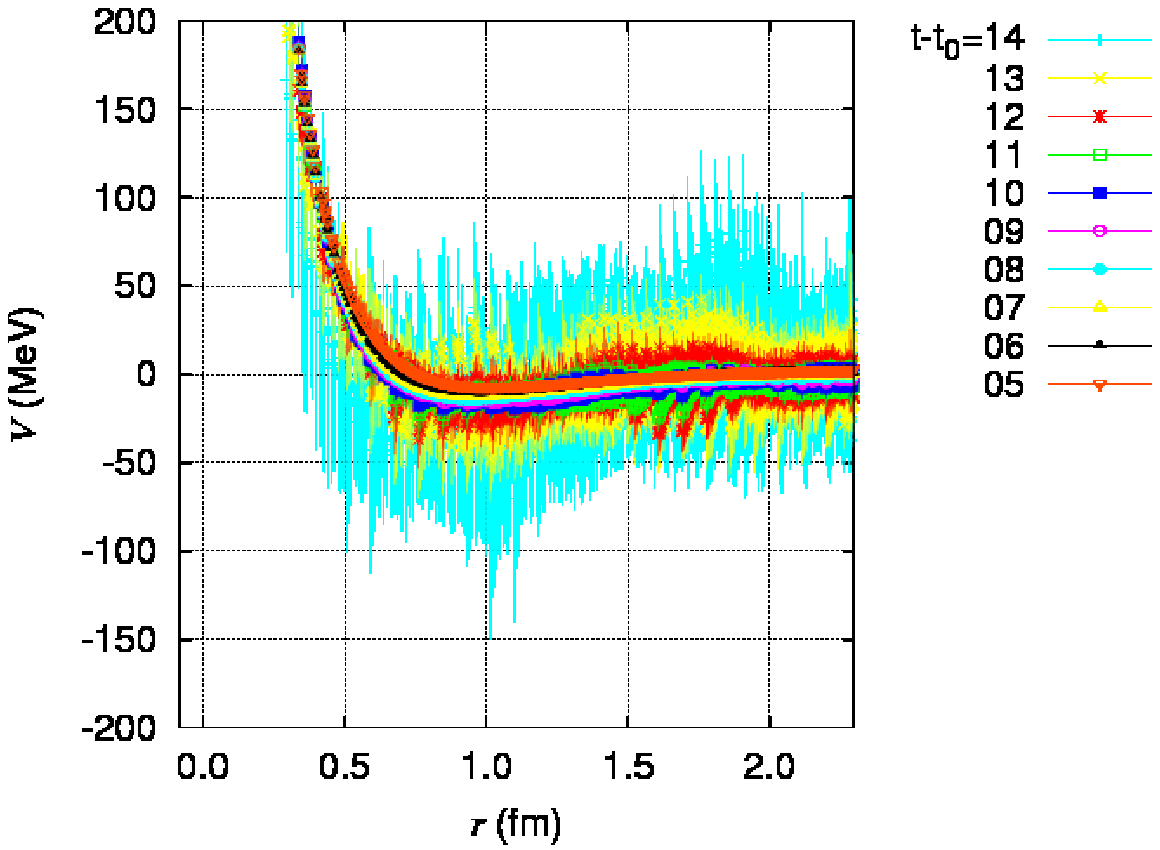}
%
 \end{minipage}~
 \hfill
 \begin{minipage}[t]{0.33\textwidth}
  \centering \leavevmode
%
  \includegraphics[width=0.99\textwidth]{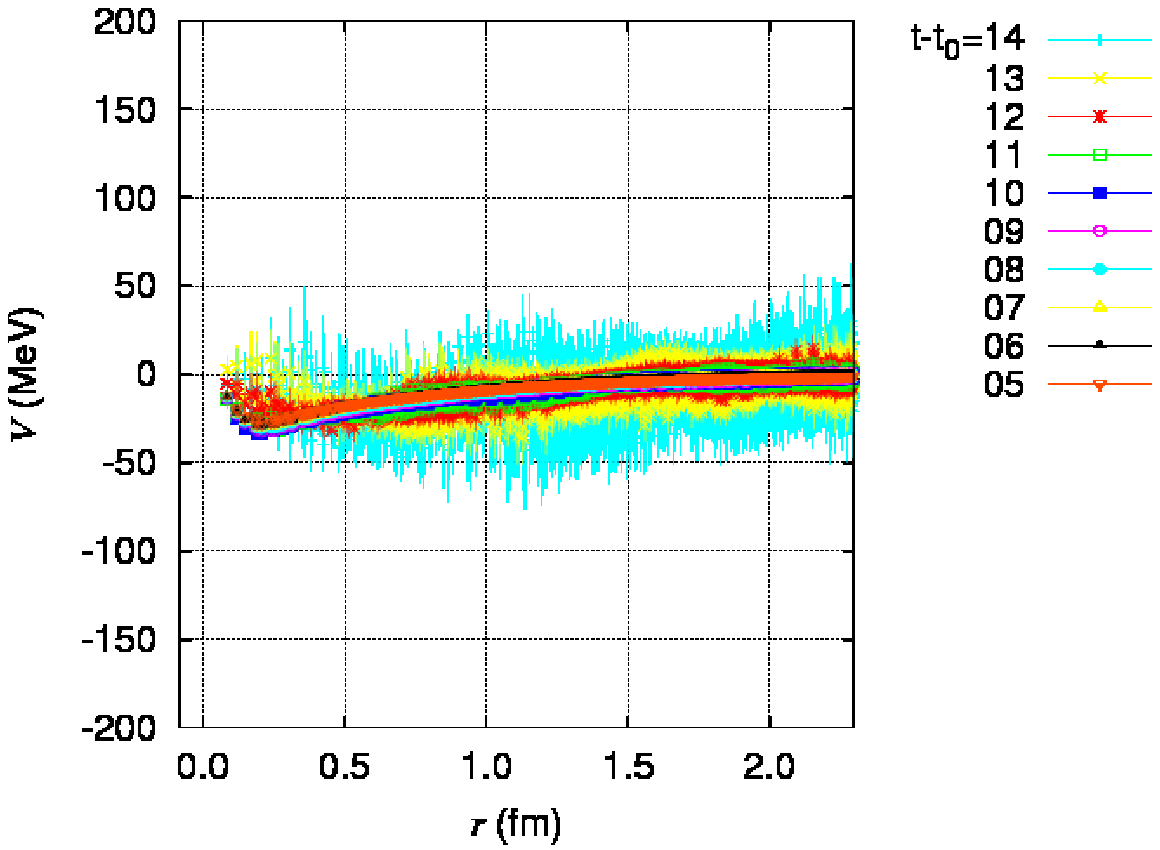}
%
 \end{minipage}
 \caption{Three $\Sigma N\!\!-\!\!\Sigma N (I\!\!=\!\!1/2)$ potentials of 
   (i) $^1S_0$ central (left), 
   (ii) $^3S_1\!-\!^3D_1$ central (center), and 
   (iii) $^3S_1\!-\!^3D_1$ tensor (right). 
   \label{VC3E1_VT3E1_VC1S0_SN_2I1}}
\end{figure}
%
%
Fig.~\ref{VC3E1_VT3E1_VC1S0_SN_2I1} shows three 
$\Sigma N-\Sigma N (I=1/2)$ diagonal potentials; 
(i) the central potential in the $^1S_0$ (left), 
(ii) the central potential in the $^3S_1-^3D_1$ (center), and 
(iii) the tensor  potential in the $^3S_1-^3D_1$ (right). %
Very strong repulsive core is seen in the $^1S_0$ central potential. 
The flavor $\bm{8}_s$ {\it irrep.} %
could %
influence 
the potential; 
we have 
$|\Sigma N\rangle ={1\over\sqrt{10}}(3|\bm{8}_{s}\rangle-|\bm{27}\rangle)$ 
in the flavor SU(3) limit. 
The statistical fluctuations 
in the strongly repulsive channel seems to be large. 
There are short range repulsive core and medium range attractive well in 
the $^3S_1-^3D_1$ central potential.

\section{Summary}

In this report, 
the study of $YN$ interactions with $S=-1$ is presented that is 
based on almost physical point lattice QCD calculation. 
The phase shifts are calculated for the $\Sigma N$ ($I=3/2$) interaction 
in both the $^1S_0$ and $^3S_1-^3D_1$ channels. 
The phase shift in the $\Sigma N$ ($I=3/2$,$^1S_0$) channel shows 
that the interaction is attractive on average. 
The phase shift $\bar{\delta}_{0}$ in the $^3S_1-^3D_1$ channel 
shows that the $\Sigma N$ ($I=3/2$,$^3S_1$) interaction is repulsive. 
These results are qualitatively consistent with recent 
studies~\cite{Fujiwara:1996qj,Arisaka:2000vu,Beane:2012ey,Haidenbauer:2013oca}. 
In the isospin $I=1/2$ channels, 
the $\Lambda N-\Sigma N$ coupled-channel potentials are presented. 
The potentials in the $^1S_0$ %
have still large 
statistical fluctuations because the number of statistics 
in the spin-singlet is factor 3 smaller than the number of statistics 
in the spin-triplet. 
In addition, the %
contribution from flavor $\bm{8}_{s}$ {\it irrep.} 
in the 
$\Sigma N$ ($I=1/2$, $^1S_0$) could break down the signal in the 
$\Sigma N$ ($I=1/2$, $^1S_0$) potential. 
Further analysis to finalize the calculations to obtain physical quantities 
are in progress and 
will be reported elsewhere. 
%


%
\section{ACKNOWLEDGMENTS}
We thank all collaborators in this project, above all, 
members of PACS Collaboration for the gauge configuration generation. 
The lattice QCD calculations have been performed 
on the K computer at RIKEN, AICS 
(hp120281, hp130023, hp140209, hp150223, hp150262, hp160211, hp170230),
HOKUSAI FX100 computer at RIKEN, Wako (
G15023, G16030, G17002)
and HA-PACS at University of Tsukuba 
(14a-25, 15a-33, 14a-20, 15a-30).
We thank ILDG/JLDG~%
which serves as an essential infrastructure in this study.
This work is supported in part by 
MEXT Grant-in-Aid for Scientific Research 
(JP16K05340, JP25105505, JP18H05236),
and SPIRE (Strategic Program for Innovative Research) Field 5 project and 
``Priority issue on Post-K computer'' (Elucidation of the Fundamental Laws
and Evolution of the Universe) and 
Joint Institute for Computational Fundamental Science (JICFuS).
%

%

%
%
%
%

\end{document}